\documentclass[iop,apjl,twocolumn,showpacs,superscriptaddress,groupedaddress]{emulateapj}  % for review and submission
\usepackage{graphicx}\usepackage{dcolumn}\usepackage{bm}\usepackage{amssymb}\usepackage{amssymb}\usepackage{amsmath}\usepackage{hyperref}
\usepackage{soul,color}

\newcommand{\boldblue}[1]{{#1}}

\begin{document}

% \title{Lagrangian Mapping Approach to Generate Intermittency and its Application in Plasma Turbulence}
% \author{P.~Subedi, R.~Chhiber, J.~A.~ Tessein,  M.~ Wan and W.~H.~ Matthaeus}
% \affiliation{Department of Physics and Astronomy, University of Delaware, Newark, DE 19716, USA}

\title{Charged particle diffusion in isotropic Random magnetic fields}
\author{P.~Subedi$^1$, W.~Sonsrettee$^{2,3}$, P.~Blasi$^{4,5}$, D.~Ruffolo$^2$, W.~H.~ Matthaeus$^1$, D.~Montgomery$^6$, P.~Chuychai$^{1}$, P.~Dmitruk$^{1,7}$, M.~Wan$^{1,8}$, T. N.~Parashar$^{1}$, R.~Chhiber$^1$}
\affiliation{
$^1$Department of Physics and Astronomy, University of Delaware, Newark, Delaware 19716, USA\\
%$^2$ New Zealand\\
%$^3$Faculty of Engineering and Technology, Panyapiwat Institute of Management, Nonthaburi 11120, Thailand\\
%$^4$Department of Physics, Faculty of Science, Mahidol University, Bangkok 10400, Thailand and
%Thailand Center of Excellence in Physics, CHE, Ministry of Education, Bangkok 10400, Thailand
%\\
%$^5$Departamento de F\'\i sica (FCEN-UBA \& IFIBA-CONICET), Buenos Aires, Argentina
%\\
%$^6$Department of Physics and Astronomy, Dartmouth College, 6127 Wilder Laboratory, Hanover, NH 03755, USA
$^2$Department of Physics, Faculty of Science, Mahidol University, Bangkok 10400, Thailand\\
$^3$Now at: Faculty of Engineering and Technology, Panyapiwat Institute of Management, Nonthaburi 11120, Thailand\\
$^4$INAF/Osservatorio Astrofisico di Arcetri, Largo E. Fermi, 5 - 50125 Firenze, ITALY
\\
$^5$Gran Sasso Science Institute (INFN), Viale F. Crispi, 7, 67100 L'Aquila, Italy\\
$^6$Department of Physics and Astronomy, Dartmouth College, Hanover, NH 03755, USA
\\
$^7$Now at: Departamento de F\'\i sica Facultad de Ciencias Exactas y Naturales, Universidad de Buenos Aires and IFIBA, CONICET Ciudad Universitaria, 1428 Buenos Aires, Argentina
\\
$^8$Now at: Department of Mechanics and Aerospace Engineering, Southern University of Science and Technology, Shenzhen, Guangdong 518055, China}

\pacs{}

\begin{abstract}

The investigation of the diffusive transport of charged particles in a turbulent magnetic field remains a subject of
considerable interest. Research has most frequently concentrated on determining the diffusion coefficient in
the presence of a mean magnetic field. Here we consider diffusion
of charged particles in fully three-dimensional
isotropic turbulent magnetic fields 
with no mean field, which may be pertinent to many astrophysical
situations. We identify different ranges of particle energy depending
upon the ratio of the Larmor radius of the charged particle to the
characteristic outer length scale of the turbulence. Two different theoretical
models are proposed to calculate the diffusion coefficient,
each applicable to a distinct range of particle energies.
The theoretical results are compared with those from computer
simulations, showing good agreement.

\end{abstract}

\maketitle

\section {Introduction}

% turbulence present in the system under study \citep{aloisio2004diffusive,casse2001transport}. 
Transport of charged particles in many astrophysical systems is governed 
by the highly turbulent magnetic field which leads to the diffusion of charged particles in space.
Some of the most notable ideas 
in the study of diffusion of charged particles
in magnetic turbulence are quasilinear theory \citep{jokipii1966cosmic}, field 
line random walk theory {\citep[FLRW; e.g.,][]{jokipii1968random},
and nonlinear guiding center theory \citep{matthaeus2003nonlinear}.
These theories focus on 
calculating the diffusion of charged
particles in directions parallel and perpendicular to a 
constant mean magnetic field ${\bf B}_0$, under influence of 
a fluctuating magnetic field ${\bf b}({\bf x})$ that depends on position
${\bf x} $ but not on time. 
A considerable 
modification to these theories is required when the 
fluctuations are isotropic with zero (or very small) mean field. 
{Such modification turns out to be of the highest importance when describing the transport of cosmic rays (CRs) 
in the Galaxy, as well 
as for the modeling of diffusive CR acceleration in astrophysical sources.}
Here we consider the problem of magnetostatic scattering 
with ${\bf B}_0 = 0$, and for fluctuations 
${\bf b}$ that are 
statistically isotropic, in terms of both polarization and spectral distribution.   

The propagation of CRs in the Galaxy is usually modeled as diffusive in a turbulent 
magnetic field where the rms fluctuations of strength $\delta b$
are of the same order of magnitude as the large 
scale field $B_0$ \citep[e.g.,][]{Jansson12}, 
$\delta b/B_0\sim 1$, although it is not clear 
whether this condition is fulfilled both in the disc and the halo of the Galaxy. On the 
other hand, supernova shocks, usually invoked to be the main sites of acceleration of 
Galactic CRs \cite[]{blasi2013}, are observed to possess intense turbulent magnetic 
fields, where the large scale field, if any, only affects the development of the fast 
growing instabilities that lead to the existence of intense turbulent fields \cite[]{caprioli2015}. 
At the shock itself, the field is probably well modeled as isotropic with a negligible mean field.

%The reversals in the orientation of background fields in {\bf the} Galaxy which are of the same order of fluctuating fields imply the existence of such isotropic turbulent fields with negligible mean field \citep{minter1996observation,candia2004diffusion}.
 
There have been some studies of diffusion of charged particles in magnetic turbulence 
without a mean field 
\citep[e.g.,][]{CasseEA01, parizot2004gzk, de2007numerical, plotnikov2011particle, snodin2016global}, 
but a clear theory covering all ranges of particle energies is still lacking. In particular, the 
simulations carried out by \cite{de2007numerical} clearly showed that in the limit of 
large $\delta b/B_0\gg 1$ the diffusion coefficient of particles with Larmor radius much smaller 
than the energy containing scale of the turbulent field closely resembles the one naively 
estimated from quasi-linear theory, a rather curious result since such theory applies to the 
opposite limit. When the field is purely turbulent, the combination of particle diffusion 
and random walk of magnetic field lines is expected to play a crucial role.
Diffusion of magnetic field lines in isotropic turbulence with zero mean field 
was examined by \cite{sonsrettee2015magnetic}, 
and that paper is 
in some ways 
an antecedent of 
the present work. We note that the high energy theory 
described below was originally presented by David Montgomery\footnote{
Presentations at the Santa Fe Workshop, Oct 5, 2004, and 
APS-Division of Plasma Physics Mini-Conference, November 15, 2004.}.

The transport of charged particles in interplanetary space and 
the interstellar medium, including in 
regions of particle acceleration, 
is highly influenced by the presence of turbulent magnetic fields
and their spectral distribution. 
The nature of particle transport in these fields 
also depends on particle energy. 
In general, higher 
energy particles, with gyro-radius larger than the correlation 
length of the magnetic field, 
will sample many uncorrelated field lines within one gyration.
Lower energy particles with gyro-radius much smaller than 
the magnetic field correlation length, on the other hand, 
will see a relatively coherent large scale field, 
and their transport will be 
heavily affected by resonances with local magnetic fluctuations.

%%%%numerics formerly here

In this paper we classify the diffusive behavior
of charged particles 
into three different regimes based on the ratio of 
the Larmor radius $R_L$ of the charged particle
to the characteristic outer length scale of turbulence $l_c$: 
a high energy region with $R_L/l_c \gg 1$, 
an intermediate energy region with $R_L/l_c \sim 1$,
and the low energy region with $R_L/l_c \ll 1$. 
Corresponding to the two extreme inequalities,
we will develop two corresponding 
theoretical approaches for  
particle diffusion in the extreme energy ranges.
First, in the high energy limit, 
the path of the particle 
experiences only small deviations from its initial trajectory 
due to small angle scattering on 
magnetic field irregularities. 
In this case velocity diffusion is achieved when the 
magnetic fluctuations probed by the particles become uncorrelated. 
%{\bf Since the high energy theory uses a straight line as an unperturbed trajectory, the theory is expected to work well %%for $R_{L}/l_{c}\gg1$, although it seems to agree reasonably well with the results of our numerical simulations down to %%$R_{L}/l_{c}\sim 0.5$.
At the opposite extreme, particles with  
very small gyro-radius
will gyrate about the local field
produced by the large scale fluctuations
while experiencing perturbations, the most effective 
of which will be at scales comparable to the 
gyroradius.
This leads to resonant interactions 
and a random walk of the particle pitch angle. 
The particle guiding center will eventually reverse direction
and parallel spatial diffusion is 
achieved. 
The range of validity of both the 
high energy
and the low energy
asymptotic theories will be extended by
building in  
additional decorrelation effects in the 
relevant Lagrangian correlation functions,
thereby providing an 
accurate description 
of the particle mean free paths also 
in the intermediate range of energies.

%\hlgreen{The intermediate energy particles, in the range $0.1\lesssim R_L/l_c\lesssim 2$,
%are neither completely tied to a field line nor are their velocities correlated 
%for a long period of time. A combination of the low energy theory and a nonlinear
%theory is used to describe the intermediate energy behavior. The nonlinear 
%thoery makes an assumption that the velocity decorrelation of particles is
%exponential. This theory also works as well as high energy theory at $R_L/l_c\gtrsim 2$
%when the decorrelation time of the velocity is long and therefore the particle 
%trajectories are straight for a long period of time.}

The paper is organized as follows: in \S \ref{sec:theory} 
we write down 
the fundamental equations describing the diffusive motion of the particles 
in velocity (or momentum) space. In \S \ref{high_energy} we specialize 
the theory to the high energy regime. In \S \ref{int_energy} we 
show that a nonlinear theory can be developed that reproduces the 
high energy behavior of the diffusion coefficient 
(\S \ref{high_energy}) for $R_{L}/l_{c}\gg 1$ while also 
describing the results of numerical simulations down to 
$R_{L}/l_{c}\approx 0.5$. Finally, in \S \ref{sec:low_energy} we describe our 
formulation of a low energy theory that, while accounting for particle-wave 
resonances, also keeps memory of perpendicular decorrelation, 
thereby describing simulation results for $R_{L}/l_{c}\lesssim 0.5$. 
A detailed comparison of our theoretically calculated spatial 
diffusion coefficients with the 
results of numerical simulations is discussed in
 \S \ref{sec:comparison}. We summarize in \S \ref{sec:concl}.
Three Appendices are included. The 
first derives the relationship between velocity diffusion and
real space diffusion for isotropic turbulence.
The second estimates the angular deflection of a field line 
over scales in the inertial range and justifies neglecting 
such deflection in the low energy theory.
The third compares the numerical simulation results with 
the theories developed in this work when applied to interplanetary parameters.
%%%%%%%%%%%%%%%%%%%%%%%%%%%%%%%%%%%%%%%%%%%%%%%%%%%%%%%%%%%%%%%%%%%%%%%%%%%%%%%%%%
%%%%%%%%%%%%%%%%%%%%%%%%%%%%%%%%%%%%%%%%%%%%%%%%%%%%%%%%%%%%%%%%%%%%%%%%%%%%%%%%%%

\section {Velocity space diffusion and spatial diffusion}\label{sec:theory}

Since the turbulence is isotropic, and the 
the particle speed $|{\mathbf{v}}|$ is constant in the absence of 
electric fields, the velocity space diffusion tensor 
is expected to have an isotropic form \citep{batchelor1953theory},

\begin{equation}
D_{ij}(v)=(\delta_{ij}-{\hat v}_i{\hat v}_j)D_v\label{iso_dv},
\end{equation}
which implies that 
\begin{equation}
D_v=\frac{1}{2}Tr[D_{ij}(v)]\label{dij_dv}.
\end{equation}

The distribution function of test particles obeys the Fokker-Planck 
equation in velocity space, which when spatial gradients are present 
can be written as
\begin{equation}
\frac{\partial f}{\partial t}+\mathbf{v}\cdot \nabla f= \frac{\partial}{\partial v_i}D_{ij}\frac{\partial f}{\partial v_j}\label{fokker_plank_gradient}. 
\end{equation} 
Plugging in the isotropic form of the velocity space diffusion (Equation \ref{iso_dv}) into Equation \ref{fokker_plank_gradient} we get,
\begin{equation}
\frac{\partial f}{\partial t}+\mathbf{v}\cdot \nabla f= D_v\nabla_{\perp}^2f,\label{fokker_plank_angular}
\end{equation}
where $\nabla_{\perp}^2$ stands for that part of the velocity space Laplacian in spherical co-ordinates that involves no
radial derivatives.

To make a connection with spatial diffusion 
we construct 
a  multiple time scale solution of Equation \ref{fokker_plank_angular},
by breaking down time and spatial scales into fast ($\tau$, $\xi$) and slow variables ($T$, $X$).
We seek a solution with $f=f^{(0)}+\epsilon f^{(1)}+\epsilon^2f^{(2)}+\dots$ where
$\epsilon \ll 1$. 
The $q^{th}$ order distribution function 
can be expanded in the spherical harmonics as:
\begin{equation}
f^{(q)}=\sum_{l,m}C_{lm}^{(q)}(\mathbf{x},v,t)Y_{lm}(\theta,\phi)\label{spher_harm}.
\end{equation}
The details of the multiple time scale solution are given in 
Appendix A where a relationship between the velocity space diffusion coefficient $D_v$
and spatial diffusion coefficient $\kappa$ is obtained:
\begin{equation}
\kappa=\kappa_{xx}=\kappa_{yy}=\kappa_{zz} = \frac{v^4}{6D_v}\label{wib_mont}.
\end{equation}
The mean free path ($\lambda$) in terms of the diffusion coefficient is 
simply
\begin{equation}
\lambda=\frac{3\kappa}{v}. \label{eqn:mean_free}
\end{equation}
%\begin{equation}
%\frac{\partial F}{\partial t} = \frac{v^4}{6D_v}\nabla_x^2F\label{rel_vel_spatial}, 
%\end{equation}
%$F(\mathbf{k},v,t)$ is isotropic in $v$.

%Hence from Equation \ref{rel_vel_spatial} we get a relation between 
%spatial diffusion coefficient $\kappa$ and the velocity space
%diffusion coefficient $D_v$
%\begin{equation}
%\kappa_{xx}=\kappa_{yy}=\kappa_{zz} = \frac{v^4}{6D_v}\label{wib_mont}
%\end{equation} 

Particle statistics may be related to the spatial 
diffusion coefficient
by the Taylor-Green-Kubo (TGK) formula
\citep{taylor1922diffusion,green1951brownian,kubo1957statistical}:  
\begin{equation}
\kappa_{xx}\equiv \lim_{t\to\infty}\frac{\langle\Delta x^2\rangle}{2t}=\int_0^\infty
dt\,\langle v_x(0)v_x(t)\rangle\label{tgk}.
\end{equation}
This formula, with $v_x$ interpreted as the guiding center velocity in the $x$-direction, was
used to calculate
the diffusion coefficient in theories that have a mean magnetic field, e.g.,
BAM theory \citep{bieber1997perpendicular}, the original NLGC theory \citep{matthaeus2003nonlinear},
etc. 

The velocity space diffusion coefficient
can also be written in a TGK formulation as:
\begin{equation}
D_{ij}(v)=\int_0^\infty dt\left\langle\left.\frac{dv_i}{dt}\right|_{0}\left.\frac{dv_j}{dt}\right|_{t}\right\rangle. \label{vel_iso}
\end{equation}
To calculate the rate of change of velocity in Equation \ref{vel_iso} we use the 
Newton-Lorentz equation of motion 
\begin{equation}
\frac{d{\bf v}}{dt}=\frac{q}{m\gamma c}({\bf v\times b})=\alpha({\bf v\times b})\label{new_lor},
\end{equation} 
where $q$ and $m$ are the particle charge and mass, respectively,
$\gamma$ is the Lorentz factor, $c$ is the speed of light, $\mathbf{b}$
is the magnetic fluctuation field and we define $\alpha=q/(m\gamma c)$.
Apply the Newton-Lorentz equation in Equation \ref{vel_iso} to find
\begin{equation}
D_{ij}(v)=\alpha^2\int_0^\infty dt\;\;\epsilon_{i\alpha\beta}\epsilon_{j\gamma\eta}\langle v_\alpha(0)v_\gamma(t)
 b_\beta(0)b_\eta[{\bf x}(t)] \rangle.
\end{equation}
Assuming that the particle velocity and magnetic field are uncorrelated
(as would be the case, e.g., 
for an isotropic particle distribution)
and that the turbulence is statistically homogeneous,
one finds
\begin{equation}
D_{ij}(v)=\alpha^2\epsilon_{i\alpha\beta}\epsilon_{j\gamma\eta}\int_0^\infty dt\langle v_\alpha(0)v_\gamma(t)\rangle
\langle b_\beta(0)b_\eta[{\bf x}(t)] \rangle\label{dij_int}.
\end{equation}
Equation \ref{iso_dv} is used to simplify Equation \ref{dij_int}.
The high energy theory and nonlinear theory differ
in the treatment of the correlation tensors in Equation \ref{dij_int} 
(more details are in Sections \ref{high_energy} and \ref{int_energy}). 
Section \ref{sec:low_energy} contains a theoretical description of 
diffusive transport for low energy particles. 

%\maketitle

\section {High Energy Theory}\label{high_energy}

Very high energy particles with Larmor radii  
much larger than the correlation scale
will experience 
only minor deflections from their original path
as they complete a distance equivalent to the
correlation length of the magnetic field.
This corresponds to 
the high energy theory developed in this section,
applicable when $R_L/l_c\gg 1$.

The appropriate simplifying assumptions
in this case are 
that the displacement follows a straight line $\mathbf{x}(t)=\mathbf{v}t$, 
and,
that 
the velocity autocorrelation is simply 
$\langle v_\alpha(0)v_\gamma(t) \rangle = v_{\alpha}v_{\gamma}$.
Using these, Equation \ref{dij_int} can be written as:

\begin{equation}
D_{ij}(v)=\alpha^2\epsilon_{i\alpha\beta}\epsilon_{j\gamma\eta}v_{\alpha}v_{\gamma}\int_0^\infty dt\;\;
\langle b_\beta(0)b_\eta(\mathbf{x}(t)=\mathbf{v}t) \rangle.\label{iso_dv3}
\end{equation}

In view of 
Equation \ref{dij_dv}, 
computing the trace of Equation \ref{iso_dv3} 
gives an expression
for the high energy 
velocity space diffusion coefficient,
\begin{equation}
D_v=\frac{\alpha^2v\delta b^2l_c}{3}\label{dv_relation},
\end{equation}
where $\delta b$ is the rms magnetic field and the correlation length is
\begin{equation}
l_c=\frac{v\int_0^{\infty}dt\langle b_i(0)b_i(vt,0,0)\rangle}{\delta b^2}. 
\end{equation}

Notice that $D_v$ is independent of the spectrum of turbulence because 
the only thing that matters is the fact that most energy is in magnetic
fluctuations at a fixed scale $\sim l_c$
and at that scale we may assume that the effective magnetic field seen by the particle
is the rms field $\delta b$. 
Plugging $D_v$ in Equation \ref{wib_mont}, 
in terms of 
gyrofrequency $\Omega_0=\alpha\delta b$, the spatial diffusion coefficient is then

\begin{equation}
\kappa_{xx}=\kappa_{yy}=\kappa_{zz}=\frac{v^3}{2\Omega_0^2l_c}.\label{high_energy_diff}
\end{equation}

This result is not entirely 
new to the astrophysics community 
\citep[see, e.g.,][]{AloisioBerezinsky04}, although we
give a more formal derivation. 
An intuitive derivation of
Equation \ref{high_energy_diff} proceeds as follows. 
Suppose that 
most of the power in the magnetic field spectrum is 
on a spatial scale $l_c$, which is much smaller
than the Larmor radius of particles.
Then one can think of 
a very large but otherwise arbitrary
distance $l_d$ as being tiled with cells of size $l_c$. In each cell,
when $R_L\gg l_c$, one has a deflection of the order $l_c/R_L\ll 1$.
The average angle of deflection when 
there are $n$ scatterings satisfies

\begin{equation}
\langle\theta^2 \rangle = n\left(\frac{l_c}{R_L}\right)^2,
\end{equation}
where $n=l_d/l_c$. Now, the average deflection angle is of order unity
when 
\begin{equation}
l_d=\frac{R_L^2}{l_c}.
\end{equation}
At that point the distance is also $l_d\sim vt$ (because the
displacement from the unperturbed trajectory is small), and
the diffusion coefficient is:
\begin{equation}
\kappa = \frac{l_d^2}{2t}=\frac{l_dv}{2}=\frac{R_L^2v}{2l_c}=\frac{v^3}{2\Omega_0^2l_c},
\end{equation}
which is identical to Equation \ref{high_energy_diff}.

\section{Nonlinear Extended High Energy Theory}\label{int_energy}

For rigidity decreasing towards unity from 
large values, 
the assumption above that 
the unperturbed particle trajectory 
is a straight line with constant velocity becomes 
less accurate. 
Instead, we assume 
the velocity autocorrelation is
exponential with a characteristic decorrelation time $\tau$,
\begin{equation}
\langle v_{\alpha}v_{\gamma}(t)\rangle=(v^2/3)\delta_{\alpha\gamma}e^{-t/\tau}.\label{vel_dec}
\end{equation}
%When $R_L/l_c\lesssim2$, one cannot make an 
%assumption that the unperturbed trajectory of the particle is a straight line
%with constant velocity. The rigidity of the particle is smaller 
%and the particle trajectories are more easily bent.
%Here a nonlinear theory is formulated based on the TGK formula where the 
%velocity is assumed to decorrelate with the characteristic 
%decorrelation time $\tau$. The theory is expected to be particularly 
%useful at the intermediate energy range where there is a 
%transition from the low energy to the high energy limit.
%The velocity autocorrelation is modeled 
%as isotropic with 
%\begin{equation}
%\langle v_{\alpha}v_{\gamma}(t)\rangle=(v^2/3)\delta_{\alpha\gamma}e^{-t/\tau}.\label{vel_dec}
%\end{equation}
However, making use of the TGK formula, Equation \ref{tgk},
one sees  that the spatial diffusion coefficient $\kappa_{xx}$ is 
related to the time scale $\tau$ by 
\begin{equation}
\kappa_{xx}=\frac{1}{3}v^2\tau.
\end{equation}
That is, from Equation \ref{wib_mont}
\begin{equation}
\tau = \frac{v^2}{2D_v}.\label{tau}
\end{equation}
Then, using Equations \ref{dij_dv}, \ref{dij_int} and
\ref{vel_dec} one finds
that
\begin{equation}
D_v=\frac{1}{2}{\rm Tr}[D_{ij}(v)]
=\alpha^2\int_0^\infty dt\frac{v^2}{3}e^{-t/\tau}\langle b_i(0)b_i[{\bf x}(t)] \rangle,\label{D_v_inc}
\end{equation}
where there is an implied summation over $i$.  
The closure for the Lagrangian 
correlation in this equation
should now be reconsidered. 
In particular, we question
the approximation of a straight-line unperturbed trajectory
$\mathbf{x}(t)=\mathbf{v}t$
that was previously used in the high energy theory.
To allow for the effect of perturbations
of this trajectory, $\mathbf{x}(t)$ may be treated as a random variable. 
Following the familiar procedure 
used in other nonlinear diffusion
theories \citep[e.g.,][]{matthaeus2003nonlinear},
Corrsin's
independence hypothesis \citep{corrsin1959atmospheric} is used to write 
the magnetic correlation in terms of the
power spectrum in Fourier space  to obtain
\begin{equation}
\langle b_i(0)b_i[{\bf x}(t)] \rangle=\int d{\bf k} P_{ii}({\bf {k}})
\langle e^{i {\bf {k}}\cdot{\bf {x}}(t)}\rangle.\label{init_RBD}
\end{equation}
%%%%%%%%%%%%%%%%%%%%%%%%%%%%%%%%%%%%%%%%%%%%%%%%%%%%%%%%%%%%%%%%%%%%%%%%%%%%%
%%%%%%%%%%%%%%%%%%%%%%%%%%%%%%%%%%%%%%%%%%%%%%%%%%%%%%%%%%%%%%%%%%%%%%%%%%%%%
In order to
evaluate the characteristic 
functional $\langle e^{i {\bf {k}}\cdot{\bf {x}}(t)}\rangle$,
we now use the approximation of a Gaussian distribution 
of displacements, along with the
random ballistic decorrelation (RBD) approximation
\citep[][]{ruffolo2012random}. 
The latter was first introduced in describing the magnetic field line random
walk by \cite{ghilea2011magnetic}.
Here, the result is 
\begin{equation}
\left\langle e^{i {\bf {k}}\cdot{\bf {x}}(t)}\right\rangle=\left\langle e^{ik\mu v t}\right\rangle_\mu=\frac{1}{2}\int_{-1}^1d\mu\; e^{ik\mu v t}=\frac{\sin(kvt)}{kvt}\label{RBD}.
\end{equation}
The random ballistic model is justified since the particles undergo ballistic motion 
at earlier times before they reach the asymptotic diffusive regime.

With $\tau$ given by Equation \ref{tau}, substituting Equation \ref{RBD} into Equation \ref{init_RBD} and then into Equation \ref{D_v_inc} gives

\begin{equation}
D_v=\alpha^2\;\frac{v^2}{3} \int dkE(k) \int_0^\infty dt\frac{\sin(kvt)}{kvt} e^{-2tD_v/v^2}\label{dv_intermediate} ,
\end{equation}
where $E(k)=4\pi k^2P_{ii}(\mathbf{k})$ is the omnidirectional energy spectrum.
If we take the limit $\tau=v^2/(2D_v)\rightarrow\infty$ in Equation \ref{dv_intermediate},
Equation \ref{dv_relation} is recovered, which is the high energy limit.
Hence, the nonlinear theory allows departures from  
the high energy limit, but also 
recovers the exact form of the high energy limit 
as $\tau$ becomes large. More generally, 
Equation \ref{dv_intermediate} is an implicit equation for $D_v$ and 
Equation \ref{wib_mont} gives the spatial diffusion coefficient.\\
It is worth pointing out that another 
standard approximation
would  be  to
treat the trajectories of the particles as having
a diffusive distribution (DD) for the purpose of calculating the
Lagrangian magnetic correlation function}
\citep{matthaeus2003nonlinear,bieber2004nonlinear,shalchi2006extended,shalchi2010unified}.
In that case, the ensemble average
on the left hand side of Equation \ref{RBD} would then be
$\left\langle e^{i {\bf {k}}\cdot{\bf {x}}(t)}\right\rangle=e^{-k^2\kappa_{xx} t}$.
This alternative approach, however, fails to predict the high energy 
behavior of the particles where the particles
are ballistic for a long time before undergoing multiple
deflections to reach the diffusive limit. For the intermediate
range DD converges with the RBD model.

\section{Low Energy Quasilinear Theory}\label{sec:low_energy}

Low energy particles ($R_L/l_c\ll 1$) 
behave entirely differently 
because they experience a local mean magnetic field 
due to large scale fluctuations.
These small gyroradius particles 
typically 
scatter before moving far enough 
for the local mean 
field to average to zero. 
In such circumstances, there are two 
dominant effects that 
are expected to 
contribute to transport and diffusion: 
Field Line Random Walk (FLRW) and a kind of 
resonant wave particle scattering in 
which the local field acts as a mean field
that organizes the particle 
gyro-motion. 

In the presence of FLRW alone, particles would follow magnetic field
lines and achieve spatial diffusion 
as the magnetic field lines diffuse in space. 
Decorrelation of the particle trajectories
due to this 
mechanism \citep{jokipii1968random, hauff2010scaling}
gives rise to an energy-independent mean free path.
This would not explain the results of our numerical simulations, as discussed below.
One may, however, estimate from 
a simple Kolmogorov turbulence theory 
the degree to which the 
field lines bend for distances shorter than a correlation scale,
as is done in Appendix B. 
The conclusion is that 
the angular deflection 
of the mean field seen by a particle in 
moving over a scale $l$
is 
small provided that 
$l/l_c$ is small, i.e., 
the scale $l$ 
lies deep in the inertial range. 
This conclusion is also empirically 
supported, in that the 
numerical results (Figure \ref{rel_diffusion})
show that at low energy 
the particle mean
free path is smaller 
than the correlation length of the
magnetic field. For lower energy the local mean field, 
due to the large scale parts of the spectrum, 
becomes increasingly coherent.

Making the assumption that the 
local mean field remains well defined 
for long enough distance, 
we may examine 
whether 
that resonant scattering of particles on fluctuations with wavenumber around the 
inverse of the Larmor radius \citep{jokipii1966cosmic,fisk1974fokker} 
provides effective scattering,
in the regime in which turbulence is weak compared with a 
locally evaluated 
large scale mean magnetic field. 
Of course if the 
small scale power is  
suppressed, then one recovers 
the FLRW result \citep[see also,][]{karimabadi1992physics, mace2012velocity}.

In the presence of a guide field, the original quasilinear 
theory \citep{jokipii1966cosmic} represents the standard 
tool for calculating resonant pitch angle diffusion coefficients of 
particles.  It  is successful in predicting the resonant scattering 
of the particles in the case of a slab fluctuation field 
(wave vectors parallel to the mean field) but may 
require some modification in other 
fluctuation geometries \citep[e.g.,][]{qin2002subdiffusive}. 
Non-linear theories 
are also useful 
in attaining 
better agreement with the results 
of numerical simulations of particle propagation 
\citep{dupree1966perturbation, dupree1967nonlinear,owens1974effects,matthaeus2003nonlinear,shalchi2004nonlinear,shalchi2006extended}.

%We have also attempted to develop a non-linear theory that works well for high and intermediate energy particles. However, when the particles are strongly tied to the field line, then the scattering of the particles along the field line is mostly dominated by the resonant scattering of the particle with the small scale fluctuations of the field where quasilinear theory works best.

In the present case of isotropic turbulence, with no guide field, 
the assumptions of QLT can no longer be applied, although 
we are encouraged to proceed based on the above 
reasoning concerning the local mean field. 
To be specific, 
it is useful to first put forward a physical explanation of the approach we 
propose to describe the low energy regime: 
if the power spectrum of magnetic 
fluctuations is such that most power is on scales of order $\sim l_{c}$, 
then particles with Larmor radius much smaller than $l_{c}$ move following a 
roughly ordered magnetic field line at least until a distance of $\sim l_{c}$ 
has been covered. 
On such scales the propagation is diffusive in the direction 
of the local magnetic field, although particles suffer little motion in the 
direction perpendicular to that of the local field. Let us refer to this parallel 
diffusion coefficient as $\kappa_{\parallel}$, although the physical meaning of 
this quantity should be kept in mind. The effective velocity of particles in the 
direction of the local field is $v_{p}\simeq \kappa_{\parallel}/l_{c}$. When particles 
move over many times the coherence scale $l_{c}$, their transport in the directions perpendicular 
to the original local field becomes evident (due to isotropic turbulence) and one can 
estimate the global diffusion coefficient as:
\begin{equation}
\kappa(p) = \frac{1}{3} L_{c} v_{p} \simeq \frac{1}{3} L_{c} \frac{\kappa_{\parallel}}{L_{c}} \simeq \frac{1}{3} \kappa_{\parallel}.
\label{eq:kappap}
\end{equation}
In other words, provided the propagation is diffusive on small scales, the global 
diffusion coefficient on large scales proceeds with a similar diffusion coefficient 
to that calculated using the local magnetic field as an {\it effective local guide field}. 
The problem is now reduced to calculating the diffusion coefficient experienced 
by particles along the local magnetic field. This can be done by 
using a close analogy to the case of QLT.

The parallel spatial diffusion coefficient $\kappa_{\parallel}$, is calculated 
using a well known relationship between $\kappa_{\parallel}$ and the pitch 
angle diffusion coefficient $D_{\mu\mu}$ \citep{jokipii1966cosmic, earl1974diffusive, hasselmann1970scattering},
\begin{equation}
\kappa_{||}=\frac{v^2}{8}\int_{-1}^{1}d\mu\;\frac{(1-\mu^2)^2}{D_{\mu\mu}}.\label{kappa_par}
\end{equation}
The diffusion coefficient along a fixed direction (e.g., the $x$-direction) is given by
\begin{equation}
\kappa_{xx}=\kappa_{\parallel}\cos^2\theta+\kappa_{\perp}\sin^2\theta,\label{k_along_axis}
\end{equation}
where $\theta$ is the angle between the magnetic field direction and the $x$-axis.
Since the particles are expected to travel parallel to the local mean field 
and undergo resonant pitch angle scattering, which eventually causes a random walk
in the parallel direction with little average perpendicular motion, we neglect
$\kappa_{\perp}$ in Equation \ref{k_along_axis} and average over all 
directions to obtain 
\begin{equation}
\kappa_{xx}=\kappa_{yy}=\kappa_{zz}=\kappa_{\parallel}/3,\label{dir_avg_diff}
\end{equation}
consistent with Equation \ref{eq:kappap}.

Using the Taylor-Green-Kubo formula, the pitch angle diffusion 
coefficient is given by

\begin{equation}
D_{\mu\mu}({\mu})=\int_0^{\infty} dt\;\langle \dot\mu\dot\mu^{\prime}  \rangle \label{8},
\end{equation}
where $\mu$ is the pitch angle at time $t=0$ and $\mu^{\prime}$ is the pitch 
angle at a later time $t$.
Particles are assumed to follow a local field line.
The initial direction of the field line can be assumed 
without loss of generality to be the 
$z$-direction and using $v_z=\mu v$, we get

$$
D_{\mu\mu}({\mu})= \frac{1}{v^2} \int_0^{\infty}dt\;\langle \dot v_z\dot v_z^{\prime}  \rangle.
$$
Using Equation \ref{new_lor} one can write
\begin{multline}
D_{\mu\mu}({\mu})=\frac{\alpha^2}{v^2}\int_0^{\infty}dt\;[\langle v_xv_x^{\prime} b_yb_y^{\prime} \rangle+\langle v_yv_y^{\prime} b_xb_x^{\prime} \rangle
\\
-\langle v_xv_y^{\prime} b_yb_x^{\prime} \rangle-\langle v_yv_x^{\prime} b_xb_y^{\prime} \rangle]\label{Dmu_exp}.
\end{multline}

The Cartesian 
$x$ and $y$ components of the velocities are given by
\begin{equation}
\begin{split}
v_x=v_{\perp}\sin\phi_0\;\;\;\;v_y=v_{\perp}\cos\phi_0\;\;\;\;\;\;\;\;\;\;\;\\
v_x^{\prime}=v_{\perp}\sin(\Omega t+\phi_0)\;\;\;\;v_y^{\prime}=v_{\perp}\cos(\Omega t+\phi_0)\label{vxvy},
\end{split}
\end{equation}
where $v_{\perp}=v\sqrt{1-\mu^2}$ is the perpendicular component of the velocity,
$\Omega$ is the local gyrofrequency and
$\phi_0$ is the initial gyrophase of the particle motion.
%$\Omega$ is the gyrofrequency of the particle given by
%$\Omega = qb/(\gamma mc)=\alpha b$, where b is the 
%magnitude of the local field that the particle is following.
Assuming the product $b_ib_j^{\prime}$ is 
independent of $\phi_0$, assuming axi-symmetry 
of magnetic fluctuations along the local mean field 
($\langle b_xb_x^{\prime}\rangle=\langle b_yb_y^{\prime} \rangle$ and 
$\langle b_xb_y^{\prime}\rangle=\langle b_yb_x^{\prime} \rangle$), using 
elementary 
identities and after averaging over the 
initial gyrophase  $\phi_0$, Equation \ref{Dmu_exp} reduces to

%\begin{equation}
%D_{\mu\mu}(\mu)=\frac{\alpha^2v_{\perp}^2}{v^2}\left[\int_0^{\infty}\cos(\Omega t)\langle b_yb_y^{\prime} \rangle + \sin(\Omega t+2\phi_0)\langle b_yb_x^{\prime} \rangle\right]\label{Dmu_int}.
%\end{equation}

\begin{equation}
D_{\mu\mu}(\mu)=\frac{\alpha^2v_{\perp}^2}{v^2}\left[\int_0^{\infty}dt\;\cos(\Omega t)\langle b_yb_y^{\prime} \rangle\right]\label{dmumu_int}.
\end{equation}
%\begin{figure}\centering
%\includegraphics[scale=1.2,angle=360]{Final_hybrid_Dmumu.pdf}
%\caption{Numerical result of the pitch angle diffusion coefficient for non-%relativistic particles
%at $R_L/l_c=0.01$. 
%} \label{running_pitch_num}
%\end{figure}
%Since the particles are following straight magnetic field lines taken to be in the z direction,
%the Fourier transform of the correlation function $R_{yy}(z)=\langle b_yb_y^{\prime}\rangle$, in terms of the 
%power spectrum $P_{yy}(\mathbf{k})$ is given by

The Lagrangian two-time correlation function
$\langle b_yb_y'\rangle = \langle b_y(0,0) b_y(\mathbf {x}(t),t)\rangle$
is greatly simplified using the QLT-like assumption that 
the particles locally follow straight 
magnetic field lines in the $z$-direction with constant pitch angle, so that 
$z=v\mu t$.
Using Corrsin's hypothesis \citep{corrsin1959atmospheric}, which is exact for 
slab fluctuations that vary only along $z$, the Fourier transform 
of the correlation function $\langle b_yb_y^{\prime}\rangle$ 
in terms of the power spectrum $P_{yy}(\mathbf{k})$ becomes
\begin{equation}
\langle b_yb_y^{\prime} \rangle=\int\int\int dk_xdk_ydk_z
P_{yy}(\mathbf{k})e^{ik_zv\mu t}\langle e^{ik_xx} e^{ik_yy}\rangle\label{one_dim_tran}.
\end{equation}
In the following two subsections, 
we proceed to evaluate 
Equation \ref{one_dim_tran} by adopting 
different 
approximations for the 
characteristic functional $\langle e^{ik_xx} e^{ik_yy}\rangle$
that describes the statistics
of the local random perpendicular displacements $x(t)$ and $y(t)$.

%%%%%%%%%%%%%%%%%%%%%%%%%%%%%%%%%%%%%%%%%%%%%%%%%%%%%%%%%%%%%%%%%

\subsection{Standard QLT Approach}\label{quaslinear}

When the perpendicular displacements 
$x(t)$, $y(t)$
are very small, we may adopt the 
approximation  
$\langle e^{ik_xx} e^{ik_yy}\rangle \approx 1$.
This can be viewed as an approximation that the particle
is at its guiding center, or that the fluctuations $b_x$ and $b_y$
are independent of $x$ and $y$.
This approximation is implemented in standard
QLT, and is 
exact in one-dimensional slab geometry \citep{jokipii1966cosmic}.
Using this
in Equation \ref{one_dim_tran}, one finds
\begin{equation}
\langle b_yb_y^{\prime} \rangle=\int dk_z\left[\int\int dk_xdk_y P_{yy}(\mathbf{k})\right]e^{ik_zv\mu t}\label{one_dim_tran2}.
\end{equation}
In the usual way 
we define $E_y(k_z)=\int\int P_{yy}(\mathbf{k})dk_xdk_y$  as 
a one dimensional reduced transverse
spectrum function.
Substituting 
Equation \ref{one_dim_tran2} in Equation \ref{dmumu_int}, and 
carrying out the time integral yields a Dirac delta function that 
defines the resonance. Then, the integral over the parallel wave number ($k_z$)
gives
\begin{equation}
D_{\mu\mu}=\frac{\pi\alpha^2(1-\mu^2)}{v}\;\left[\frac{E_y\left(k_z=\frac{\Omega}{v|\mu|}\right)}{|\mu|}\right],\label{pangle_quas}
\end{equation}
which is the standard QLT result.

One modification applicable to isotropic turbulence 
that we would like to discuss in detail is related to 
the $\mu$ dependence of $D_{\mu\mu}$ in Equation \ref{pangle_quas}.  
It is well known that quasilinear theory does not provide 
correct pitch angle diffusion coefficients for pitch angles 
close to $90^\circ$ ($\mu\approx0$) where nonlinear effects 
are important \citep{bieber1988cosmic,tautz2008solving,shalchi2009nonlinear}.  
This problem was discovered in the years after quasilinear 
theory had been proposed \citep{jones1973new, owens1974effects,  goldstein1976nonlinear}.  
The strict quasilinear calculation using Equation \ref{pangle_quas}
(for typical spectra of turbulence) has $D_{\mu\mu}\to0$ as $\mu\to0$, 
which is applicable in slab turbulence \citep{qin2009pitchangle}
where nonlinear effects are weak. In most other geometries, 
however, for any realistic finite amplitude fluctuations, 
nonlinear orbit effects \citep{karimabadi1992physics} allow 
particles to scatter more easily through the neighborhood 
of $\mu = 0$ than would be expected from QLT calculations 
\citep[see also][]{qin2009pitchangle}. Our numerical results also indicate 
that the quasilinear theory for isotropic turbulence as 
derived using Equations \ref{pangle_quas}, 
\ref{kappa_par} and \ref{dir_avg_diff} gives mean 
free paths much larger than those obtained from numerical 
simulation, since the parallel diffusion coefficient is 
unrealistically amplified by $D_{\mu\mu}$ near zero in the denominator.

To account for such nonlinear effects, we consider an ansatz 
that the pitch angle dependence of $D_{\mu\mu}$ is of the 
form $D_{\mu\mu}\propto1-\mu^2$ and set $|\mu|=1$ inside the square 
bracket of Equation \ref{pangle_quas} so that $D_{\mu\mu}$  has a finite 
nonzero value at $\mu = 0$.  Before implementing this, we checked 
our numerical simulation results for the pitch angle distribution.  
For $D_{\mu\mu}\propto1-\mu^2$ (isotropic scattering) the 
eigenfunctions of the diffusion operator are Legendre polynomials, 
and at late times when the pitch angle distribution is nearly isotropic, 
the distribution should be dominated by the most slowly evolving 
eigenfunctions, $P_0=1$ and $P_1=\mu$, yielding a nearly linear 
pitch angle distribution.  We indeed found this in our simulation 
data.  Thus we implemented the ansatz, which is more appropriate 
for our case of isotropic turbulence with $B_0 = 0$ than for a perturbative 
situation with $b \ll B_0$ where nonlinear effects are weak.  The 
ansatz is also justified a posteriori in section \ref{sec:comparison} 
where the theoretical predictions now provide a much better fit to the numerical results.
   
With this physically motivated modification, the simplified 
pitch angle diffusion coefficient can be written as
\begin{equation}
D_{\mu\mu}=\frac{\pi\alpha^2(1-\mu^2)}{v}\;E_z\left(k_z=\frac{\Omega}{v}\right).\label{dmu_quas}
\end{equation}
If we assume that the local mean field is constant throughout the 
system with magnitude  
$B_{local}$ then in Kolmogorov turbulence $E_z\sim k_z^{-5/3}$ and (from Equation \ref{new_lor}) 
$\alpha=\Omega/B_{local}$ with $\Omega = v/R_L$, it is straightforward to show that
the substitution of Equation \ref{dmu_quas} into \ref{kappa_par} gives 
$\kappa\sim vl_c\left({R_L}/{l_c}\right)^{1/3}$ and the mean free path
(Equation \ref{eqn:mean_free}) scales as 
$\lambda\sim l_c\left({R_L}/{l_c}\right)^{1/3}$.

A further correction 
that we apply here
is related to the variability 
of the magnetic field strength
in an isotropic random field. 
In the realization of turbulence 
used here,
the components of the magnetic field have a Gaussian distribution
(this is also usually a reasonable approximation for 
fully developed turbulence).
Therefore the gyrofrequency 
$\Omega = qb/(\gamma mc)=\alpha b$ and Larmor radius 
$R_L=\gamma mvc/(qb)=v/(\alpha b)$
vary in space as the magnetic field strength $b$ changes.
In fact $b$ is likely to undergo major changes during the 
parallel scattering process described by $\kappa_{||}$.
It is fairly simple to show that
$b$ has a Maxwellian distribution given by 
\begin{equation}
\mathcal{F}(b)db=3^{3/2}\sqrt{\frac{2}{\pi}}\frac{b^2}{\delta b^3}e^{-\frac{3b^2}{2\delta b^2}}db,\label{Maxwellian}
\end{equation}
where $\delta b$ is the root mean square field strength.
With the above Maxwellian distribution, 
we use Equation \ref{dmu_quas} to compute the average $\overline{D_{\mu\mu}(\mu)}$ 
over the local field $b$
before substitution in Equation \ref{kappa_par} to obtain
\begin{equation}
\kappa_{||}=\frac{v^2}{8}\int_{-1}^{1}d\mu\;\frac{(1-\mu^2)^2}{\overline{D_{\mu\mu}}}.\label{kappa_par_avg}
\end{equation}
Equation \ref{dir_avg_diff} gives the spatial 
diffusion coefficient along a particular axis. The mean free path scaling of 
$\lambda\sim l_c\left({R_L}/{l_c}\right)^{1/3}$ is also maintained.

\subsection{Extended Low Energy Theory}\label{quaslinear_extended}
As an extension to the above 
quasilinear approach, 
we now take into account the 
perpendicular displacements
that enter 
into consideration in Equation \ref{one_dim_tran}.
Assuming a Gaussian 
distribution of 
the perpendicular displacements $x$ and $y$
implies that  
\begin{equation}
\langle e^{ik_xx} e^{ik_yy} \rangle = 
e^{-\frac{1}{2}[\langle x^2\rangle k_x^2+\langle y^2\rangle k_y^2]}.\label{trans_xy}
\end{equation} 
Since we are assuming locally an unperturbed 
orbit about a well defined local mean magnetic field, 
it is clear that $\langle x^2\rangle = \langle y^2 \rangle \sim R_L^2$.
In particular, 
for pitch angle 
$\theta$ and 
gyrophase $\phi$,
\begin{equation}
x = R_L\sin\theta\cos\phi,\;\;\; y = R_L\sin\theta\sin\phi.
\end{equation}
After omnidirectional averaging of $x^2$ and $y^2$ over 
$\theta$ and $\phi$, Equation \ref{trans_xy}
can be written as
\begin{equation}
\langle e^{ik_xx} e^{ik_yy} \rangle = e^{-k_{\perp}^2R_L^2/6},\label{rl_trans}
\end{equation}
where $k_{\perp}^2=k_x^2+k_y^2$. This statistical description of the perpendicular
displacement is consistent with our earlier assumption that $b_ib_j^{\prime}$ is
independent of the initial gyrophase. 
Using Equation \ref{dmumu_int} along with Equations \ref{one_dim_tran}
and \ref{rl_trans} one finds
\begin{equation}
\begin{split}
D_{\mu\mu}(\mu)=\frac{2\pi^2\alpha^2(1-\mu^2)}{v}\;\;\;\;\;\;\;\;\;\;\;\;\;\;\;\;\\
\times\int_0^{\infty}k_{\perp}dk_{\perp}\left[
\frac{P_{yy}\left(k_{\perp},k_z=\frac{\Omega}{v|\mu|}\right)}{|\mu|}\right]e^{-k_{\perp}^2R_L^2/6}.
\label{pitch_angle_diff_int}
\end{split}
\end{equation}

According to the arguments used in section \ref{quaslinear}, we 
take into account the presence of nonlinear orbit effects 
near $90^\circ$ pitch angle and 
modify the pitch angle dependence 
so that $D_{\mu\mu}\sim (1-\mu^2)$ 
and set  $|\mu|=1$ inside the
square bracket of Equation \ref{pitch_angle_diff_int}.
For the extended case, 
the pitch angle diffusion coefficient is now given by
\begin{equation}
\begin{split}
D_{\mu\mu}(\mu)=\frac{2\pi^2\alpha^2(1-\mu^2)}{v}\;\;\;\;\;\;\;\;\;\;\;\;\;\;\;\;\\
\times\int_0^{\infty}k_{\perp}dk_{\perp}
P_{yy}\left(k_{\perp},k_z=\frac{\Omega}{v}\right)e^{-k_{\perp}^2R_L^2/6}.
\label{pitch_angle_diff}
\end{split}
\end{equation}
Note that when the exponential term is set to unity
(e.g., for $R_L \to 0$), 
this reduces to Equation \ref{dmu_quas}. 
As described in 
section \ref{quaslinear}
we average Equation \ref{pitch_angle_diff} over a Maxwellian
distribution of the magnetic field magnitude $b$
before substitution in Equation \ref{kappa_par}
to obtain our final result for 
the spatial diffusion coefficient.

The exponential term can be seen as another modification 
to the original quasilinear theory appropriate for the pitch angle
diffusion of low energy particles in isotropic turbulence.
Hence, we use this extended low energy theory in the following 
section \ref{sec:comparison} to compare with numerical simulation results.

\maketitle

\section {Comparison with Numerical Simulation}\label{sec:comparison}

\begin{figure}\centering
\includegraphics[scale=0.46,angle=360]{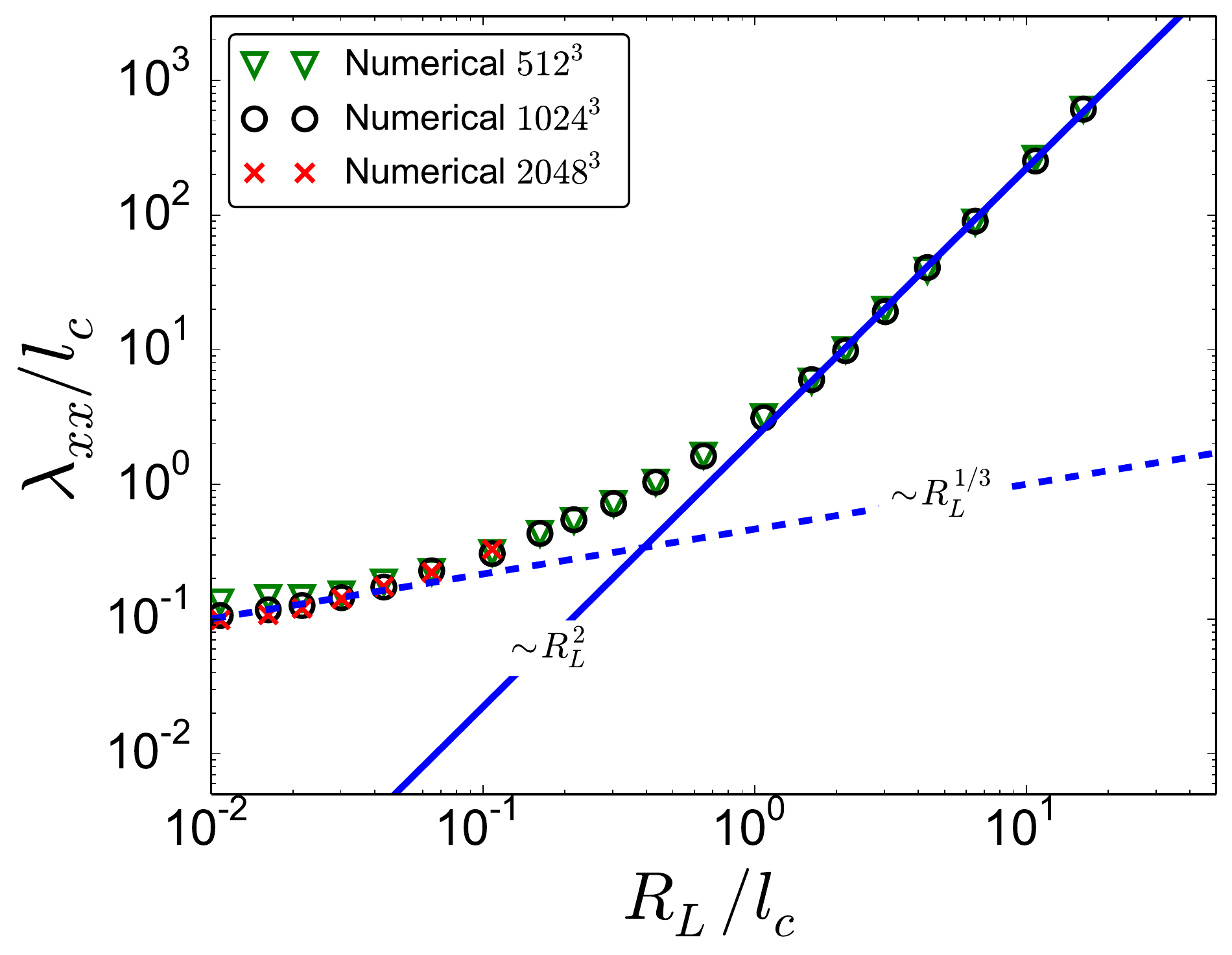}
\caption{
Mean free path of protons as a 
function of $R_L/l_c$ (gyroradius divided by 
correlation scale). The solid line is the high energy 
scaling $\lambda_{xx}\sim R_L^{2}$ and the dashed line
is the low energy scaling $\lambda_{xx}\sim R_L^{1/3}$.
The inverted triangles are the results of $512^3$ numerical simulation, circles 
are those of $1024^3$ simulation and the crosses are the results of $2048^3$
numerical simulation. \label{scaling}
}
\end{figure}

\begin{figure}\centering
\includegraphics[scale=0.46,angle=360]{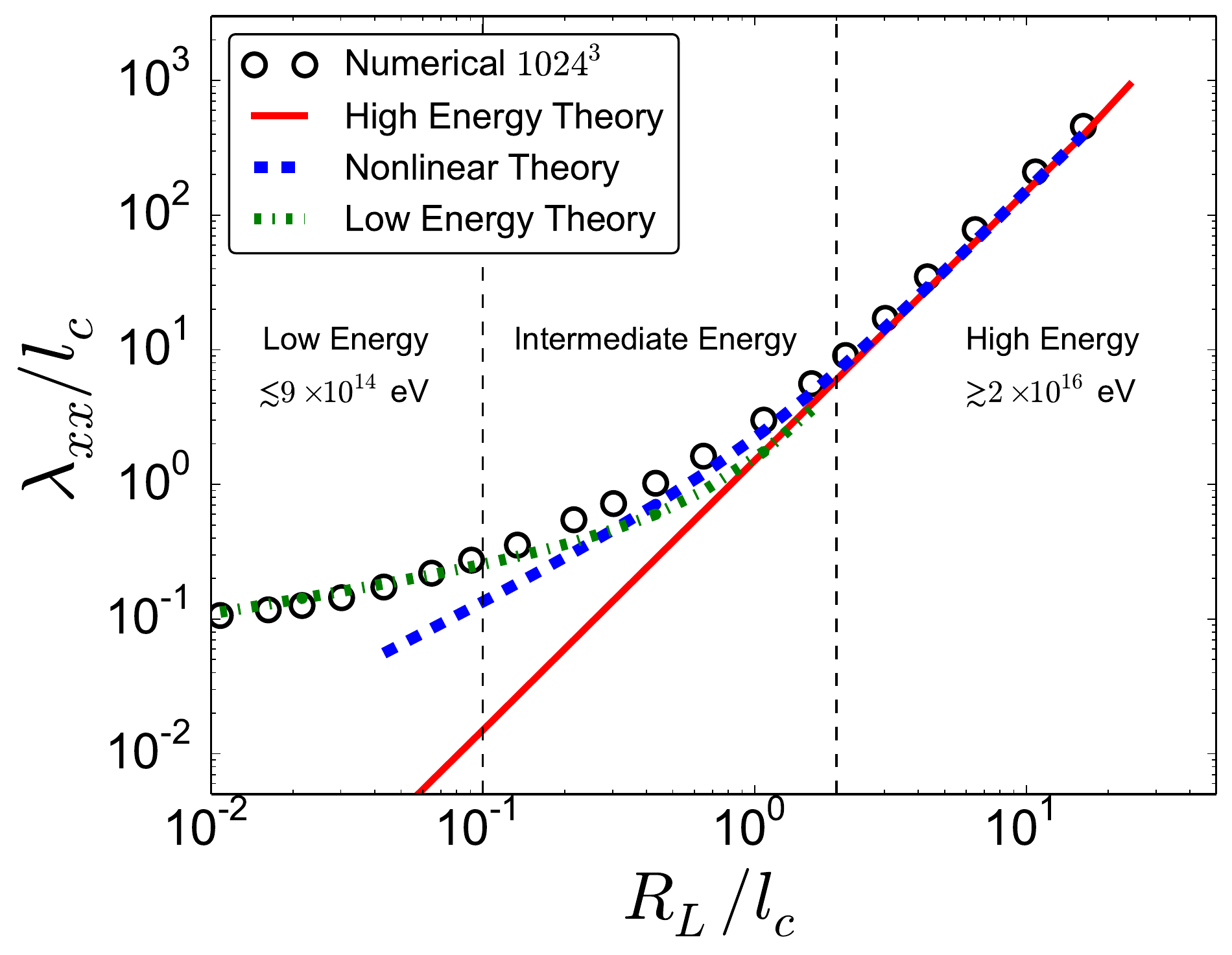}
\caption{
Theoretical vs.\ numerical results for 
cosmic rays in isotropic turbulence with zero mean field, 
showing good agreement. The circles represent the 
numerical results, solid line represents the high energy theory,
dashed line is the nonlinear theory and dotted dashed line 
is the theoretical estimate for low energies. The energy ranges are 
shown for cosmic rays in our Galaxy.
} \label{rel_diffusion}
\end{figure}

%\begin{figure}\centering
%\includegraphics[scale=0.46,angle=360]{non_rel.pdf}
%\caption{\prada{
%Theoretical vs. numerical results for nonrelativistic  
%particles in isotropic turbulence with zero mean field. 
%The circles represent the
%numerical results, solid line represents the high energy theory,
%dashed line is the nonlinear theory and dotted dashed line 
%is the theoretical estimate for low energies. The results apply to 
%any nonrelativistic particles of a given $R_L/l_c$; for 
%convenience energy ranges are shown for protons and magnetic field parameters
%for interplanetary turbulence.}
%} \label{nonrel_diffusion}
%\end{figure}

In this section we present 
results of numerical simulations of charged 
particle propagation in synthetic magnetostatic
turbulence with a specified spectrum. 
The 
numerical results are compared with the 
theoretical formulations described above. 

The numerical simulations 
make use of 
a homogeneous and isotropic magnetic fluctuation field,
generated on a spatial grid 
with a specified energy spectrum.
Trajectories of 2000 particles are obtained by numerical solution
of the
Newton-Lorentz equation, using a fifth-order Runge-Kutta method
with adaptive time-stepping. 
To satisfy the magnetostatic assumption, the velocity $v$ of the particles is chosen so that 
$v\gg v_A$, where $v_A$ is the Alfv\'en speed. The electric field is ignored
as it is of order
$v_AB/c$, where $B$ is the magnetic field and 
$c$ is the speed of light.

The random magnetic field realization is generated in a periodic box 
as described by \cite{sonsrettee2015magnetic}.
The functional form of the omni-directional spectrum used 
in the numerical simulation is given by
$$E(k)=C\lambda_c\frac{(k\lambda_c)^4}{[1+(k\lambda_c)^2]^{17/6}},$$
with a normalization constant $C$, used to control the magnetic
field strength $b$. The parameter
$\lambda_c$ is the bendover scale, which is of the same order as the 
correlation length.
This form of $E(k)$ is chosen so that 
$E(k)\propto k^4$ for low $k$ to
be consistent with strict homogeneity \citep{batchelor1969computation}, and
$E(k)\propto k^{-5/3}$ for high $k$ to represent Kolmogorov scaling in an
inertial range of turbulence.
We have chosen the Kolmogorov spectrum, as is often assumed in scattering
theory \citep[][]{harari2002astro, parizot2004gzk}, but emphasize that the 
theoretical approach can be applied to any reasonable spectrum.

%The omnidirectional 
%magnetic fluctuation energy spectrum $E(k)$
%is chosen to be a function $E(k)\propto k^4$ for low $k$ to be 
%consistent with strict homogeneity \citep{batchelor1969computation}, and $E(k)\propto k^{-5/3}$
%for high $k$ to represent Kolmogorov scaling in an inertial range of turbulence. 

The diffusion coefficient
is calculated from the asymptotic rate of increase of the mean square displacement of 
the particles
\begin{equation}
\kappa_{xx}=\lim_{t\to\infty}\frac{\langle\Delta x^2\rangle}{2\Delta t}.
\end{equation}

The results of the numerical simulations are shown in 
Figure \ref{scaling}.
The 
general asymptotic trends for the mean free path at 
low energies ($R_{L}\ll l_{c}$) and high energies ($R_{L}\gg l_{c}$) are easy 
to identify.
In the low energy 
regime the mean free path scales as $\lambda_{xx}\propto R_{L}^{1/3}$, while in the 
high energy regime $\lambda_{xx}\propto R_{L}^{2}$, and these scalings are valid 
whether the particles are relativistic or non-relativistic. 
As discussed above, 
this high energy 
scaling is obtained irrespective of the power spectrum, 
provided most power 
is concentrated on scales around $\sim l_{c}$. 
These results confirm previous functional forms proposed 
by \cite{AloisioBerezinsky04} and \cite{parizot2004gzk} and 
the numerical results of \cite{CasseEA01}, \cite{de2007numerical} and \cite{snodin2016global}.
Observations of isotopic composition \citep{obermeier2012apj,aguilar2016prl}
also support a scaling of
$\lambda_{xx}\propto R_{L}^{1/3}$ at $R_L \ll l_c$.

The resonant nature of particle scattering in the low energy regime makes 
numerical simulation rather challenging, in that the resulting mean free 
path is sensetive to the resolution 
in real space which is
equivalent to the number of 
independent degrees of freedom 
in $k$-space used to represent the 
magnetic power spectrum \citep[see also][]{snodin2016global}. 
In Figure \ref{scaling} we show the results of our 
simulations for $512^{3}$, $1024^{3}$, and $2048^{3}$ real space grid
points, from which it is 
possible to see that the lower resolution ($512^3$) case does not match correctly the 
scaling of the mean free path with Larmor radius. This is because the $512^3$
simulation does not numerically resolve the resonant scales 
corresponding to the Larmor radius of the particles. On the other hand, no 
appreciable difference can be seen in the two higher resolution cases with 
$1024^3$ and $2048^3$ simulations. 
Hence, we will use the $1024^{3}$ realization
for subsequent discussion of the 
results.

The mean free path calculated using the theoretical framework developed above 
is compared with the results of simulations in Figure \ref{rel_diffusion}.
These results may be considered as representative of propagation of 
very high energy CR protons in the Galactic magnetic field \citep{ruzmaikin2013magnetic}
with correlation length $l_c=10$ pc and root mean square magnetic field $\delta b=0.1$ nT. 
%The mean free path from theory is compared with numerical
%simulation for relativistic high energy particles in Figure \ref{rel_diffusion}.
%The physical paramters choosen for our numerical sumulation are those of 
%protons in a Galactic magnetic field \citep{ruzmaikin2013magnetic} with correlation length
%$l_c=10$ pc and root mean square magnetic field $\delta b=0.1$ nT. 
%%Similar parameters have been used in previous numerical studies 
%%(\cite{AloisioBerezinsky04,parizot2004gzk}), mostly applicable
%%to the extragalactic space where the regular magnetic field is expected to 
%%be negligible and the fluctuations are also expected not to be more than a few tens of
%%nano Gauss (\cite{kronberg1994extragalactic}, \cite{blasi1999cosmological}). 
%%The low energy particles have energy approximately less than $3\times 10^{16}$ ev,
%%and the high energy particles have energy approximately greater that $5\times 10^{17}$ ev.
The Larmor radius of the particles equals the correlation length of the magnetic field 
fluctuations at energy $E\sim9\times 10^{15}$ eV, somewhat above the knee in the all-particle 
spectrum of Galactic CRs. 
However, we note that when plotted this way, 
as mean free path vs. Larmor radius, with both quantities 
normalized to the correlation scale of the turbulence,
the actual curves are identical for any energy range. 
In Appendix C we provide a short account 
of the results of our analysis when applied to 
non-relativistic and moderately relativistic particles
in isotropic turbulence that has parameters akin to heliospheric parameters.
 
%The numerical results agree well with the theoretical 
%predictions at high and low energies. The nonlinear theory provides the 
%best prediction at the transition between low and high energies, but the 
%agreement weakens closer to the lower energy regime. 
%
%Similarly, we compare the 
%mean free paths from theoretical results and numerical
%simulations for physical parameters closely resembling interplanetary space of the solar system,
%in Figure \ref{nonrel_diffusion}.
%To estimate the energy scale for protons, 
%the correlation length is chosen to be $l_c=0.01$ AU and the root mean square magnetic field 
%to be $5$ nT \citep{burlaga2013magnetic}, for which
%$R_L=l_c$ at 1.5 GeV.
%The agreement is very similar to that for relativistic high energy particles
%(Figure \ref{rel_diffusion}). This is why we think that 
%the result can be generalized to any system as long as the turbulence
%is isotropic.

\section {Discussion and Conclusions}\label{sec:concl}

The spatial 
diffusion of charged particles in the presence of 
turbulent magnetic fields 
is significant in describing the transport of charged particles
in the interplanetary, interstellar and intergalactic media.
Most of the standard description of diffusion involves the existence of 
a non-negligible ordered field.
In this paper we 
considered the case of a vanishingly small background ordered
magnetic field and studied
the diffusive transport of charged particles in isotropic 
magnetic field (magnetostatic) 
turbulence.
The turbulent Galactic magnetic field,
where the mean magnetic field is of the order of fluctuations,
is found to have reversals in the field orientation implying the 
existence of regions with negligible regular fields \citep{minter1996observation}. 
Similarly the extragalactic voids might
have negligible regular fields \citep{kronberg1994extragalactic, blasi1999cosmological}.
\boldblue{In the highly turbulent plasma near supernova shocks that 
are actively accelerating cosmic 
rays \citep{blasi2013} there may be regions in which the 
fluctuations might be so large that
the present fully isotropic model is applicable. 
The isotropic assumption is also relevant
in other highly disturbed and turbulent plasmas, such as 
planetary wakes \citep{gombosi1979plasma}.}

\boldblue{Naturally, in the Galactic system 
the turbulent magnetic field is not the only complication
that appears in a realistic scenario of cosmic ray transport. 
The complex composition of background 
plasma and the presence of neutral hydrogen atoms that suppress
the propagation of waves through ion-neutral damping are
examples of factors that could influence the
charged particle propagation. Including these complexities
in our theoretical calculations is beyond the scope of the 
current paper. Here we have developed a systematic derivation
of charged particle diffusion in the presence of an isotropic turbulent magnetic
field and tested its validity using Monte-Carlo simulation
of particles in a synthetically generated turbulent magnetic field.
In fact, we would like to point out that incorporating the several factors 
mentioned above will probably require MHD, hybrid or PIC simulations 
that provide a self-consistent picture of the development of the cosmic 
ray transport phenomenon \citep[e.g.,][]{reville2008transport,caprioli2014sim}.}                       
	
The transport of charged particles in a magnetic field is conceptually
simple, requiring mostly the integration of particle 
trajectories using the Lorentz force (Equation \ref{new_lor}). The
detailed treatment is complex, influenced by several
parameters like the particle energy, correlation length of 
turbulence, the geometry of fluctuations, 
the presence of a dissipation range, and 
the magnetic Reynolds number.
In particular the study of transport of particles 
in isotropic magnetic field turbulence is found to 
be complicated 
by the different conditions seen by 
charged particles 
at high, low and intermediate energies as defined.
Here the controlling parameter
is the rigidity, or 
ratio 
of Larmor radius ($R_L$) to the turbulence 
correlation length ($l_c$). 
When the ratio $R_L/l_c$
is much smaller than one we refer to the low energy limit, when
it is greater than one this is the high energy limit and 
in between low and high energy lies the intermediate energy region.
We have devised two different 
theoretical models in an attempt to understand the transport processes
in these three regions.
Rather than just using parametric and scaling arguments, 
we provide systematic and thorough theoretical constructs 
to describe the diffusive transport of particles in each range of energy.

The paths of the higher energy particles are almost straight up to the 
correlation length of the magnetic field, where the particles 
undergo only a slight change from the original trajectory.
In contrast, the lower energy particles are almost strictly tied 
to their initial field lines and are pitch angle scattered due to resonance
with small scale irregularities of the field line. The intermediate 
energy particles have a more complicated nonlinear diffusion
that connects the two extremes. 

The theoretical ideas presented here 
have been tested against 
results of detailed numerical experiments using Monte Carlo
simulations of particle propagation in stochastic magnetic fields. 
The magnetic field constructed for numerical purposes has three dimensional isotropic 
fluctuations with a standard Kolmogorov spectrum and no mean field. 
Diffusion coefficients are calculated using the
microscopic displacements of particles along the trajectory.
%Since our study is more applicable to galactic systems, we present 
%the results of numerical propagation of relativistic high energy galactic 
%cosmic rays with energies above about $10^{14}$ eV. Interestingly for low energies ($R_L/l_c\ll 1$), 
%there are data available based on the confinement time of cosmic rays in the galaxy.
The diffusion coefficients obtained using our numerical simulation are compared
with theoretical predictions yielding very good agreement.

Referring again to Figure \ref{rel_diffusion},
we can see that
both the asymptotic low and high energy limits 
are fitted well by our theoretical approach. 
The high energy theory fails                                                
in the intermediate and low energy regime because 
the particle velocity changes over a shorter
period of time, contrary to the high energy assumption.
The extended low energy theory 
compares well 
with the numerical data even in the intermediate 
energy regime when $R_L/l_c \lesssim 0.5$. However, it fails when $R_L/l_c\gtrsim0.5$ 
as a consequence of the fact that the effective guide field becomes ill-defined.
The nonlinear theory best describes
the scattering of
the intermediate energy particles in the range $R_L/l_c\gtrsim0.5$.
The two extended theories 
give equal results at $R_L/l_c \approx 0.3$ where they each 
differ from the numerical results by about 30\%. 

In summary, we have devised theoretical
descriptions of charged particles in isotropic turbulence with no mean field
that are applicable to 
distinct ranges of particle energies.
Different reasoning enters in the various ranges of rigidity $R_L/l_c$.
Properly normalized, 
the results apply 
equally well to relativistic and non-relativistic particle transport,
provided that the 
random magnetic field is isotropic with zero mean.

%We thank\dots
\begin{acknowledgments}
This research is supported in part by the NSF
Solar Terrestrial Program Grant AGS-1063439 and SHINE grant
AGS-1156094, the NASA Heliophysics Grand Challenge 
Research Theory \& Modeling
Program, NNX14AI63G, the MMS
Theory and Modeling project, the Solar Probe Plus project,
POR Calabria FSE-2007/2013, EU Turboplasmas project, and by
grant RTA5980003 from the Thailand Research Fund.

\end{acknowledgments}

\section*{Appendix A: Relation between diffusion coefficients 
in velocity space and position space}\label{App1}

We proceed 
using 
spatial and temporal variables divided into 
slow variables $X$ and $T$, 
and fast variables $\xi$ and $\tau$, respectively.
The slow and the fast variables can be written in 
terms of a small parameter $\epsilon$
as:
$$
T = \epsilon^2t,\;\;\;\;X=\epsilon x
$$ 
$$
\tau = t,\;\;\;\;\;\xi = x\;\;.
$$

The motivation for different time ordering when compared to the spatial ordering
comes from the fact that, if we let $\epsilon$ be the ratio of scattering length 
scale to the transport length scale, then from the spatial diffusion 
equation a simple dimensional analysis shows that the ratio of scattering time
scale to the transport time scale should be of the order 
$\epsilon^2$ \citep[see, e.g.,][]{frisch1995turbulence}.

The derivatives can then be written as
$$
\frac{\partial}{\partial t} = \frac{\partial }{\partial \tau} + \epsilon^2\frac{\partial}{\partial T},  
$$
$$
\nabla_x = \nabla_\xi + \epsilon \nabla_X.  
$$

Considering
Equation \ref{fokker_plank_gradient} (Fokker-Plank equation), 
we seek a solution of the distribution function $f$ with 
\begin{equation}
f=f^{(0)}+\epsilon f^{(1)}+\epsilon^2 f^{(2)}+\dots\label{f_expanded},
\end{equation}
where each 
order of $f$ can be expanded in terms of the spherical harmonics 
$Y_{lm}(\theta,\phi)$ (Equation \ref{spher_harm}),
which 
satisfy
\begin{equation}
v^2\nabla_{\perp}^2Y_{lm}(\theta,\phi) = \left[-l(l+1)\right]Y_{lm}(\theta,\phi)\label{eigen_val}.
\end{equation}

The expansion
in Equation \ref{f_expanded} is inserted 
into Equation \ref{fokker_plank_angular}.
With the help of Equations \ref{spher_harm} and \ref{eigen_val} we get separate equations for each 
of the different orders of $f$.

The $O(\epsilon^0)$ terms give:
\begin{equation}
\frac{\partial f^{(0)}}{\partial \tau}+\mathbf{v}\cdot\nabla_\xi f^{(0)} = D_v\nabla_{\perp}^2f^{(0)}.\label{oe0}
\end{equation} 
Now we introduce an operator $\langle\cdot\cdot\cdot\rangle$ which is the space-time 
average over the fast variables. Since all variables
will be assumed to have finite variations in time and space, the
averaging operator gives a zero result when operating on any
quantity that may be written as a derivative with respect to fast variables. This is 
also known as the solvability condition so that we have a closed form of 
equations. Also, let $\langle f^{(q)} \rangle = F^{(q)}$.

Equation \ref{oe0} averaged over the fast variables gives:
\begin{equation}
D_v\nabla_{\perp}^2F^{(0)} = 0\label{oe1},
\end{equation}
which implies that $F^{(0)}$ is isotropic in $\mathbf{v}$.

Averaged over the fast variables, the $O(\epsilon^1)$ term is 
\begin{equation}
\mathbf{v}\cdot\nabla_XF^{(0)}=D_v\nabla_{\perp}^2F^{(1)}\label{oe2}.
\end{equation}
Without loss of generality we may pick the direction of the gradient 
to be the $z$-direction.
The left hand side of Equation \ref{oe2} has a projection only onto the $Y_{10}$
term of the spherical harmonics. The right hand side must also have a
projection onto the $Y_{10}$ term. With the help of Equations 
\ref{spher_harm} and \ref{eigen_val} in Equation \ref{oe2}, one finds  
\begin{equation}
\mathbf{v}\cdot\nabla_XF^{(0)}=-2\frac{D_v}{v^2}F^{(1)}\label{oe2_sec}\;\;\Rightarrow\;\; F^{(1)}=\frac{-v^2}{2D_v}\mathbf{v}\cdot\nabla_XF^{(0)}.
\end{equation}

When averaged over the fast variables the $O(\epsilon^2)$ term gives:
\begin{equation}
\frac{\partial F^{(0)}}{\partial T} + \mathbf{v}\cdot\nabla_X F^{(1)} = D_v\nabla_{\perp}^2F^{(2)}\label{oe3}.
\end{equation} 
Substituting the relation between $F^{(1)}$ and $F^{(0)}$ from 
Equation \ref{oe2_sec} into Equation \ref{oe3} one gets
%\begin{equation}
%\frac{\partial F^{(0)}}{\partial T} -\frac{v^2}{2D_v} \mathbf{v}\cdot\nabla_X (\mathbf{v}\cdot\nabla_XF^{(0)}) = D_v\nabla_{\perp}^2F^{(2)}\label{oe3b}.
%\end{equation}
\begin{equation}
\frac{\partial F^{(0)}}{\partial T} -\frac{v^2}{2D_v}\left(v_iv_j\nabla_{X_i}\nabla_{X_j}\right) F^{(0)} = D_v\nabla_{\perp}^2F^{(2)}\label{oe3b},
\end{equation}
where a repeated index indicates summation.  To study the diffusion 
of the bulk distribution, we consider the 
average over all directions.  From Equation \ref{eigen_val}, 
we see that the right hand side of Equation \ref{oe3b} averages to zero.  Using
\begin{equation}
[v_iv_j]_{\text{direction\;averaged}}\; = \frac{v^2}{3}\delta_{ij},
\end{equation}
we then obtain
\begin{equation}
\frac{\partial F^{(0)}}{\partial T}=\frac{v^4}{6D_v}\nabla_X^2F^{(0)}.
\end{equation}
%where a repeated index implies summation. 
%The left hand side of Equation \ref{oe3b} has a projection onto $Y_{20}$,
% but it cannot have a projection onto $Y_{00}$ since the right 
%hand side of Equation \ref{oe3b} has zero projection onto $Y_{00}$.
%The latter conclusion may be inferred from Equation \ref{eigen_val}. 
%This implies that the left hand side of Equation \ref{oe3b} must vanish
%when averaged over all the directions since $Y_{20}$ vanishes 
%when averaged over directions.
%Using the fact that in the isotropic scenario
%\begin{equation}
%[v_iv_j]_{\text{direction\;averaged}}\; = \frac{v^2}{3}\delta_{ij},
%\end{equation}
%averaging over the directions on the left hand side of Equation \ref{oe3b} 
%and setting it to zero yields:
%\begin{equation}
%\frac{\partial F^{(0)}}{\partial T}=\frac{v^4}{6D_v}\nabla_X^2F^{(0)}.
%\end{equation}
This is a diffusion equation in configuration space with diffusion coefficient 
$$
\kappa = \frac{v^4}{6D_v}.
$$

\section*{Appendix B: Angular deflection of a magnetic field line over an inertial 
range of turbulence}\label{App2}

We examine
the angular deflection of a field line over a separation $l$ from the 
perspective of turbulence theory.
The magnetic field vector and its associated 
field lines experience a change of 
direction due to the turbulence. 
If initially we 
start with no transverse component, then the transverse component at a 
separation $l$ in the inertial range of turbulence
can be estimated approximately for an ensemble of
field lines by Kolmogorov's law. 

Let $S_2^t(l)$ be the second order transverse structure function with separation $l$.
Kolmogorov's $2/3$ law states that 
$${S_2^t(l)}=\delta b_l^2=\langle[b_t(\mathbf{x})-b_t(\mathbf{x}+{l})]^2 \rangle\propto l^{2/3}. $$

If $S_2^t(l_c)$ is the second order transverse structure function at a separation 
of a correlation length $l_c$, then one finds
$$
\frac{\delta b_l^2}{\delta b_c^2}=\left(\frac{l}{l_c}\right)^{2/3}.
$$

Initially there is no transverse component and if
approximately $2/3$ of the energy is present in the 
transverse component 
at the correlation length, then
$$
\delta b_l^2\approx\frac{2}{3}\delta b^2\left(\frac{l}{l_c}\right)^{2/3},
$$
and 
$$
\frac{\delta b_l}{|\delta b|}\approx0.8\left(\frac{l}{l_c}\right)^{1/3}.
$$

If $\theta$ is the angular deflection at the mean free path of a particle  
$\lambda$, then  
$$\sin\theta=\frac{\delta b_l}{|\delta b|}\approx0.8\left(\frac{\lambda}{l_c}\right)^{1/3}.$$
For $R_L/l_c = 0.01$, which is relevant for a low energy that is numerically attainable,
our numerical results give $\lambda=0.1\;l_c$, implying $\theta\approx 20^0$.
At lower particle 
energy the relevant
angular deflection is still
smaller.
We neglect this deflection 
in the low energy quasilinear 
theory presented in Section \ref{sec:low_energy}.

\section*{Appendix C: Non-Relativistic Example}\label{App3}

\begin{figure}
\includegraphics[scale=0.46,angle=360]{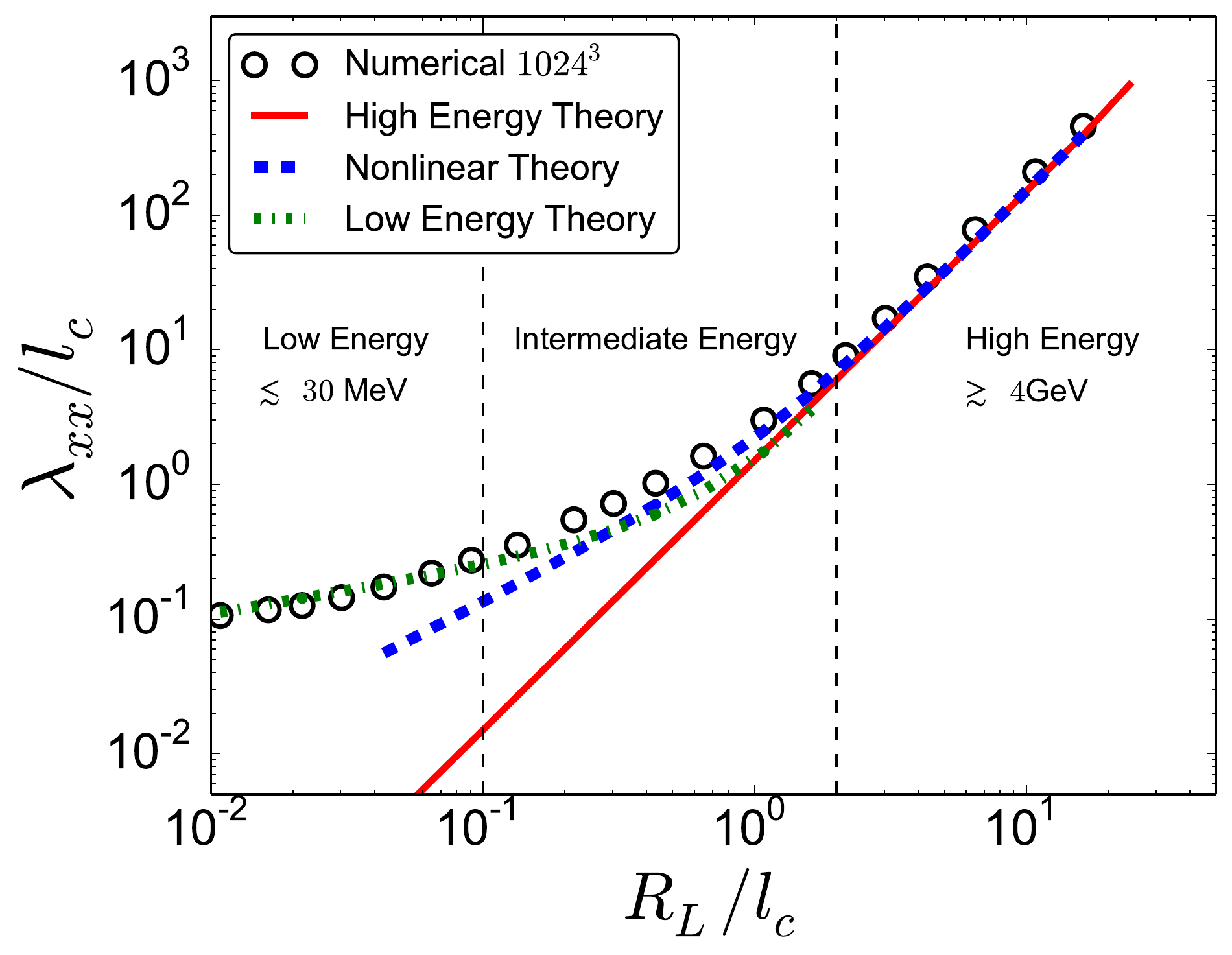}
\caption{
Theoretical vs.\ numerical results for 
non-relativistic or moderately relativistic charged particles in isotropic 
turbulence with zero mean field for
parameters as listed in Appendix C.
The circles represent the numerical results, solid line 
represents the high energy
theory, dashed line is the nonlinear theory and the dotted dashed
line is the theoretical estimate for low energies. The
energy ranges are shown for protons and magnetic field parameters
for interplanetary turbulence. The results also apply
to any non-relativistic particles of a given $R_L /l_c$.
} 
\label{fig:nonrel}
\end{figure}

To provide a broader context, in Figure \ref{fig:nonrel}
we evaluate the theory for the case of non-relativistic 
particles or moderately relativistic particles.  
For convenience we adopt parameters that 
closely resemble interplanetary space in our solar system, 
except that here no mean magnetic field is included.  
For this example, the particles are non-relativistic 
at low energies with $R_L/l_c <1$, but become moderately relativistic 
at higher energies. The correlation 
length is chosen to be $l_c = 0.01$ AU and the root mean 
square magnetic field to be 5 nT \citep{burlaga2013magnetic}. 
Then $R_L = l_c$ at 1.5 GeV. The theoretical mean free paths 
depend only on rigidity and test particle 
simulation is in close agreement whether the particles are 
relativistic or non-relativistic.

It is well known that the interplanetary turbulent magnetic field is 
represented by an anisotropic spectral model in which the fluctuations 
are of the same order of magnitude as the mean field. Numerous realistic 
studies have also been performed and compared with observational results 
in the past \citep[e.g.,][]{tautz2013diffusion}, so our study with 
isotropic turbulence and no background field should be interpreted 
as simply a test case.

\end{document}